# The Making Of A Theory Of The Vitreous Solid State: "From 1 milli-K To 1 kilo-K" ()

Giancarlo Jug[1,2]

[1]INFN (Istituto Nazionale di Fisica Nucleare)-Sezione di Pavia, Via Bassi 6, 27100 Pavia (Italy);
[2]DiSAT, Università dell'Insubria, Via Valleggio 11, 22100 Como (Italy)

## ABSTRACT

This is a short account of the basic principles of a comprehensive theory of the vitreous state, looking at glasses and their eventual "melting" into a liquid state (the inverse glass "transition") from the perspective of their low-temperature properties, then working all the way up to the melting temperature (and back again). The theory is still in the making, so only essential guidelines will be provided. There appears to be no ideal-glass transition, no ideal-glass state, but a first theoretical microscopic justification of the Vogel-Fulcher-Tammann (VFT) law can be provided, making glasses truly fascinating quantum as well as topological substances.

## 1. INTRODUCTION

It would be a redundant exercise - in this Symposium attended by Industrialists, Material Scientists, Physical Chemists and Condensed-Matter Physicists (as well as by many more distinguished Specialists, we are sure) - trying to stress the enormous importance of amorphous (glasses and thin/thick films) materials in human civilization and from its cradle's all the way to present-day's technical and technological applications. The latter range from construction to strategic electronics and biotechnology applications; as a single example we could mention the paramount importance of the amorphous tunneling barrier in superconducting Josephson-junction based quantum computing devices. Yet for physicists, particularly the condensed-matter physicists, glasses still remain somewhat of a mystery as they escape the desire to describe mathematically their physical properties somewhat independently of composition and henceforth predict new wondrous effects. For the condensed-matter theorists, glasses – though known to mankind since pre-history - still pose a formidable intellectual challenge. The crucial issue is the nature of glasses' (and amorphous film's) atomic structure at the intermediate scales, for it is known that locally the structure resembles that of the corresponding crystal forms while globally (at least at large scales) these substances hardly differ in their structure from their liquid state. It is at the intermediate scales that the fundamental degrees of freedom make their appearance in determining the physical properties of glasses and amorphous films. In this context must we then interpret P.W. Anderson's 1995 Science Magazine statement, according to which "The Deepest and Most Interesting Unsolved Problem in Solid State Theory is Probably the Nature of Glass and of the Glass Transition" [«Through the Glass Lightly», *Science* **267**, 1615 (March 17, 1995)]. Glasses, the glass "transition" and amorphous films and their properties still baffle the physics community, owing to the lack of a unifying working mathematical framework.

Most, if not all approaches to investigate the nature of glass and of the glass "transition" take their steps from the liquid and more specifically the super-cooled liquid state, where the atomic players can (presumably) still be thought to remain in thermodynamic equilibrium. It is the realm of computer simulations, like Molecular Dynamics (MD), which have been pushed to the limits to unravel the structure, properties and the nature of the "transition". MD simulations, however useful as they are, still proceed to investigate only nanoseconds of the dynamics of small systems of particles interacting through fictitious potentials. Likewise, liquid-state based theories (such as mode-coupling theory) view the problem of glass in terms of a dramatic, but unexplained, increase of the relaxation times in approaching the glass "transition". In our opinion, the hopes to solve the big problems as have been put by P.W. Anderson, and from the liquid state are as limited as trying to explain the properties of (crystalline) semiconductors and magnets starting from the study of their molten state. Not surprisingly, liquidists looking at glasses view them as "arrested liquids"; which is very useful, but "arrested" by what? Saying that glasses' molecules get to be "caged" by their neighbors' slowdown is like stating that – in traffic theory – vehicles get to be trapped by the surrounding vehicles, which is not the real cause of the traffic jam since everyone wants to move on.

At the other end of the temperature spectrum, however and fortunately, are the so-called cryogenic (temperatures $T$ less than 4 K) anomalies of glasses, which have been explained (with O.L. Anderson and H.E. Bömmel (1955) somewhat preceding P.W. Anderson et al. (1972), in our view) in terms of tunneling states (popularly known as two-level systems). In this cryogenic regime glasses and $a$-films differ considerably from their crystalline counterparts, showing up a wide range of new physical behaviors. There, the physicists had a good season in the 1970s and 1980s (particularly thanks to the then cheap price of liquid-He) and recently – with the quantum information age wanting to be delivered – this type of glassy-state physics has regained considerable interest and importance. However, the nature of the tunneling systems has never been unraveled, ultimately, and the common belief is that soon beyond 1 K the cryogenic degrees of freedom get to be destroyed by thermal vibrations and leave the stage for yet-misunderstood "bosons" responsible for the Boson-Peak behavior (anomalous quantized-vibrations type physics). The question remains, however, whether cryogenic anomalies, Boson-Peak and glass "transition" physics are somewhat related in terms of the very same main effective degrees of freedom involved. Though some scant attempts in the literature do exist, this type of question still remains totally avoided – with the notable exception of P.G. Wolynes and collaborators.

In this paper we will try to convince the reader that – at least in principle – some hope to connect all the different issues associated with the physics of glasses does exist and this thanks to the discovery – some 15 to 20 years ago – of unexpected magnetic effects in the cryogenic anomalies of nominally non-magnetic glasses. When properly handled and explained, these new magnetic effects discovered **just above 1 mK** provide a formidable new set of clues about the nature of the fundamental players in the physics of glasses, and this starting from the situation that is created in approaching the melting temperature $T_m$ (or - which is the same - the crystallization temperature $T_x$, which for silicate glasses approaches and **exceeds the 1 kK mark**). As $T_x$ is approached, even with a fast cooling rate κ (sometimes indicated also as $R$), inevitably some solid-like cluster seeds begin to form – spontaneously or activated by dirt or bad glass-forming components – for the simple reason that when cooling down the atomic players energetically prefer to organize themselves with some embryonic local degree of spatial order. It is then a matter of kinetics whether true microcrystals form and later grow to form a domain-ridden poly-crystalline solid, or whether partially-, quasi-ordered lumps grow to a limited extent and then get to be jammed one against the other, still without having the chance to form true micro-crystals (save a few isolated ones), but coalescing to a macroscopic poly-cluster (a kind of percolation phenomenon) that is nonetheless characterized by some true rigidity. This scenario is consistent with the dynamical heterogeneities (DH) scenario that has been discovered and is recognized to exist in the super-cooled liquid state of the glass-forming material, the size ξ of the solid-like (slower-particle) regions of the DH being determined by the cooling rate κ.

So it is in the super-cooled liquid state that we begin to see the true players, the fundamental degrees of freedom for the understanding of the physics of glasses (at all temperatures). The odd situation we have at present is, that below $T_g$ the glass-science community (at least in the West) that accepts the DH scenario above $T_g$ now turns its head away from a more natural heterogeneous but frozen scenario. Thus preferring homogeneous network- and metallic-glass models like the Zachariasen-Warren and Bernal models, while at the same time acknowledging partial crystallization as a rule in the polymeric glasses (plastics). While representing useful "simple" models, the said homogeneous glass scenarios (let's talk about the "perfect, or ideal glass" scenario) are at odds with the DH scenario and – as we will show below – with many important cryogenic anomalies reported in recent years. As well as with some modern theories for the Boson-Peak phenomenology. Choosing a poly-cluster, DH-turned-SH (static heterogeneities) scenario with the size ξ still growing, but not diverging, below $T_g$ in its stead, we were able to construct a mathematical model for glass structure that allows for the explanation of all the cryogenic anomalies (also when varying composition and in the presence of a magnetic field). This model (which we term the cellular model and/or the extended tunneling model (ETM)) holds the promise of resolving the decades-long Boson-Peak problem, the glass "formation" and melting problems (they are indeed separate problems) and many more issues in the grand arena of the physics of glasses. We will touch only briefly on these issues, much of the work done at non-cryogenic temperatures being largely unpublished, aiming at a description with these new ideas for the chief issue within the glass-forming liquids science: the spectacular divergence of the viscosity η of the liquid/super-cooled-liquid state approaching $T_g$ from above.

The key novelty of this approach is that one must begin understanding the solid glassy state (for it is a solid - of a new type - rather than an arrested liquid) from the lowest temperatures and then keep warming up: not the other way round, starting from the high temperature liquid and cooling down. Trying to understand glasses from the liquid state only would be, as already stated, like trying to understand metals, semiconductors, magnets, superconductors and insulating dielectrics from the physics of their molten counterparts (it would be impossible).

## 2. THE SUPERCOOLED STATE AND ITS QUASI-STATIC COUNTERPART BELOW $T_G$ AND FURTHER

We begin to review, briefly, what happens in the super-cooled state (a long-lived metastable state, as we know it to be) in terms of intermediate-range structure. So all issues of rigidity-theory (though very useful indeed for the Chemists) will be avoided. In the DH scenario, the super-cooled liquid is characterized by dynamically-rearranging regions (or clusters) called slow-particle regions and by fast-particle regions. They have a size (diameter) ξ (as well as a size distribution, which we ignore here) and they are continuously changing, almost so that if a particle jumps off its own cluster of slow particles somewhere, somewhere else another particle from the fast particle regions attaches itself to another slow cluster. It's happening everywhere and one gets the impression that – unlike in the true equilibrium liquid state – an infinite dynamical particle correlation length characterizes this situation. Mode-coupling (MC) theory predicts this to happen below a critical temperature $T_c$ (also called $T_{MC}$) where in MC-theory the viscosity η diverges as a power-law in $|T-T_{MC}|$ with some negative exponent (this in never observed, in practice), making the interval $[0,T_{MC}]$ a line of critical points. More modern theories acknowledge that something happens at $T_{MC}$ (the onset of an infinite dynamical correlation length) but that this state persists down to a second critical temperature known as the Kauzmann temperature $T_K < T_g$ (making $T_g$ somewhat ill-defined). This is what is predicted by the RFOT (random first-order theory) which views glass formation (as well as melting) as a second-order phase transition, at least for some "ideal glass" situation. Oddly, because in order to obtain the vitreous state a fast quench is necessary, otherwise a crystal is always obtained when cooling down the liquid via a reversible (quasi-static) thermodynamic transformation. In terms of bulk transport parameters, RFOT predicts ($D$ is the diffusion constant of a small Brownian particle in the medium, τ the typical relaxation time):

$$\eta, \tau, D^{-1} \propto e^{\frac{A}{S_c}}$$

Here *A>0* is a constant pertinent to the transport parameter at hand and $S_c$ is the entropy per slow-particle "cluster" which some separate theory (e.g. the Adam-Gibbs theory of CRC (continuously-rearranging clusters)) may predict to vanish linearly at $T_K$ as some (positive) constant times $|T-T_K|$. So this approach (still somewhat unsatisfying, though) leads to the VFT-law (*B* >0):

$$\eta, \tau, D^{-1} \propto e^{\frac{B}{T-T_K}}$$

which is much better than MC, but not always observed experimentally, unless *B* is temperature-dependent. Apart from this (important) issue, the DH scenario (though no-one has ever observed the dynamical heterogeneous clusters in an experiment so far, but see Section 7) offers nice, inspiring computer-generated pictures of what the super-cooled state looks like. Fig.1 shows some of the earliest pictures obtained by MD simulations and essentially all further studies lead to similar situations. Though a movie renders the situation in a much better way, essentially there are slow-particle clusters swimming in a medium of much faster-particles (the clusters being highly correlated).

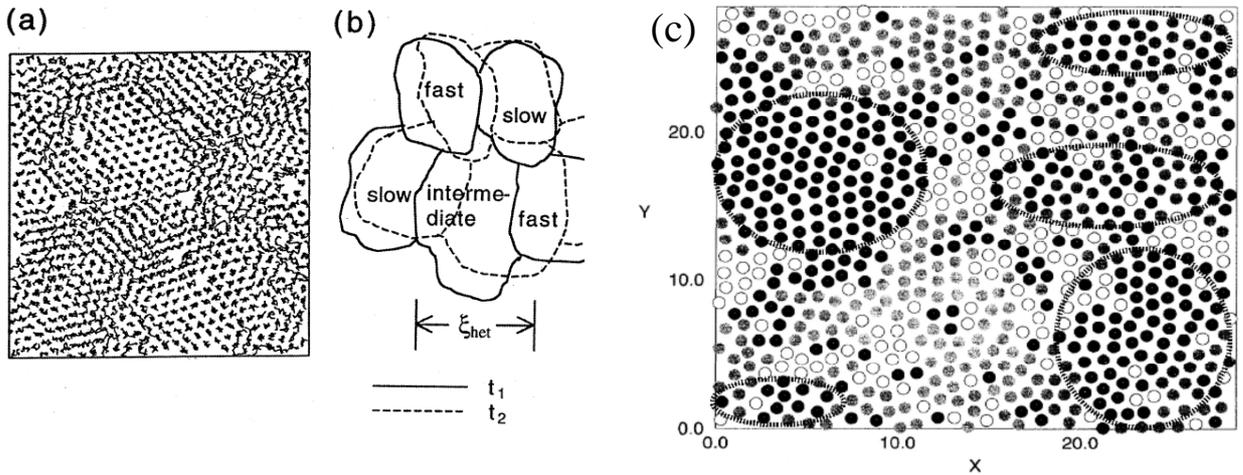

Figure 1- The (computer-simulated) supercooled-liquid heterogeneous particle structure. One early simulation (a) and its schematization (b), then to the right (c) another simulation and the (by us) ringed clusters of slow-moving particles. All for 2D Lennard-Jones like computer models (see [1] and references therein).

The main point of Fig.1 is the observation (made by the present Author first in 2004, but also by H. Tanaka and his group) that the slow-particle clusters are also better ordered; a careful mathematical description can be found in Tanaka's work [2], also for three-dimensional (3D) models. The clusters can in fact be characterized by some quasi-order parameter *q* (hexatic for 2D, icosahedral for 3D) which is close to zero outside the clusters. Fig.2 portrays a typical picture from the simulations of Tanaka et al. (which are for colloidal particles just above the glass "transition") and next to that, in Fig.3, is a cartoon of what we can conclude for the structure of the super-cooled liquid "just below" $T_g$ and then at cryogenic temperatures.

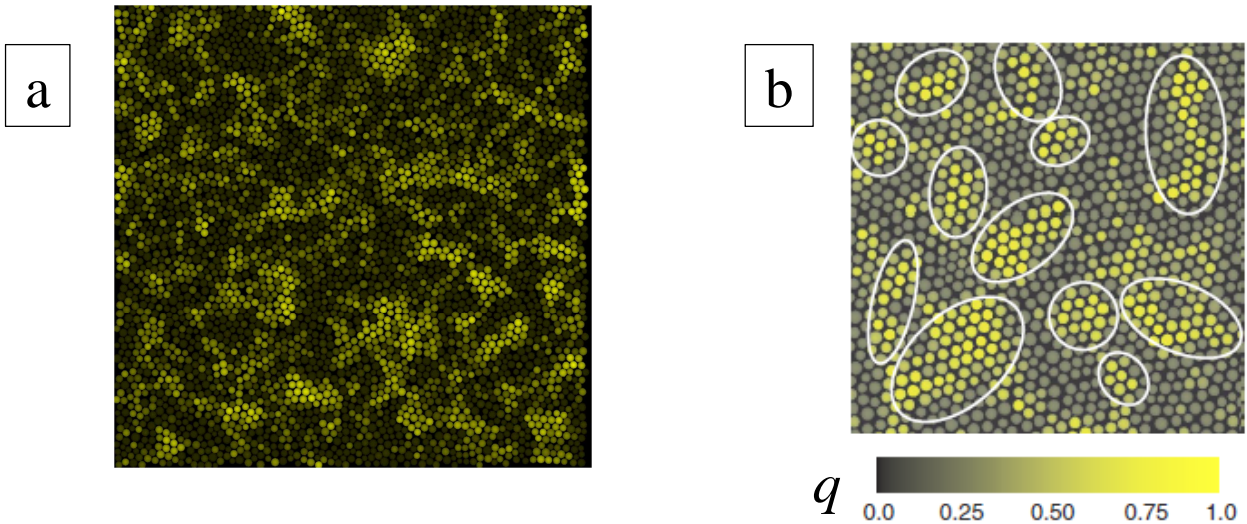

Figure 2- Simulation snapshots (in 2D, for colloidal suspensions) by the Tanaka group. In (a) a snapshot of a simulation movie, in (b) the slow-particle clusters of another snapshot are highlighted and the values of the (hexatic) quasi-order parameter $q$ reported. [2]

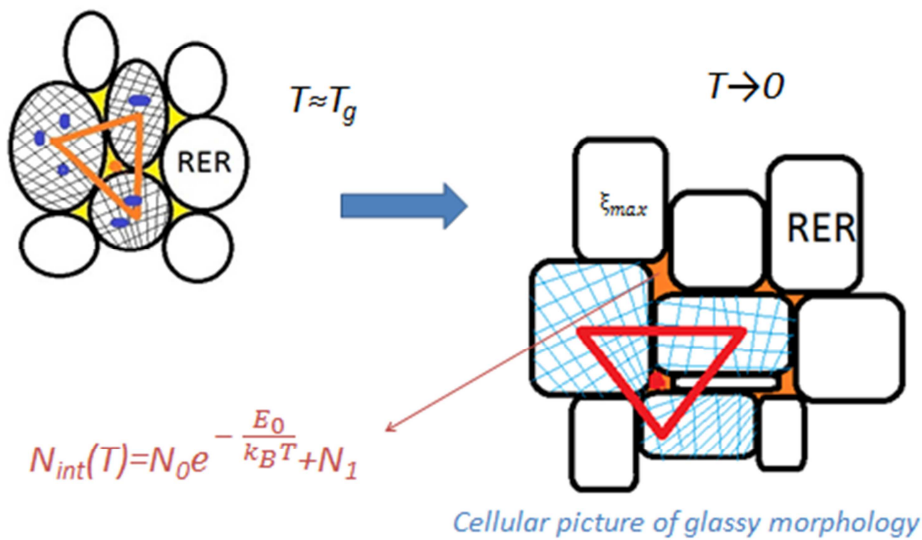

Figure 3- A cartoon (in 2D) of what should happen, inspired by Fig.2, just below the glass transformation temperature region and then at much lower, cryogenic temperatures. The triangle etc. are explained below, the formula gives the temperature-dependence of the RER (regions of enhanced regularity) "voids" particle population. [3]

So what is going on, is that as $T_g$ (or there about) is reached from above, the size ξ of the slow-particle clusters (now called RER, regions of enhanced regularity) grows without diverging till the RER jam together in a complex fashion and a solid begins to form. The process of solid formation continues, the RER re-arranging themselves in a narrow temperature range till when, at $T_K$, the jamming process stops (and the entropy per particle $S_c$ of the solid sticks to that of the crystalline counterpart). Further down with cooling, there is further consolidation of the RER shapes (not really the growth of its linear size ξ) which can grow in terms of its particle content at the expense of the still liquid-like particles that sit in the "voids" between the jammed, close-packed RER. In fact, what we learn from the DH phenomenology, is that the RER form in the super-cooled state as solid-like clusters sitting in a bath of liquid-like faster-moving particles. This phenomenology has no reason to disappear below $T_g$ and so the now jammed RER will be surrounded, in their closed-packed structure, by liquid-like material all the way to the lowest temperatures.

In a nutshell, in this cellular-model of glass these substances are nothing but amorphous opals, with (we shall see) very tiny "spheres" making up the solid, which is just a little more complicated than a **failed poly-crystalline solid** (or, in the words of A.S. Bakai, a **poly-clusterine solid**) [3] filled with remnant liquid-like matter in the RER-RER pores.

Are glasses really made in this way? Electron microscopy (HRTEM, high-resolution transmission electron microscopy) has indeed given us some (too few!) images that confirm this structure, for some thin-flakes of multi-component glassy materials. In Fig.4 we show the case for silica, $a$-SiO$_2$, amorphous $(Li_2O)_{0.5}(SiO_2)_{0.5}$ and then for the amorphous $(B_2O_3)_{0.75}(PbO)_{0.25}$ systems. The fluid-like medium in the "voids" does not show up in the images, but we are indeed dealing with $a$-opals. We can estimate the average size ξ of the RER in these picture, and this is 60 nm for the lithium-silicate and 50 nm for the lead-borate. For silica, $a$-SiO$_2$, below is also good HRTEM pictorial evidence that ξ≈10 nm, but other HRTEM pictures in the literature give cell sizes of up to 50 nm [3].

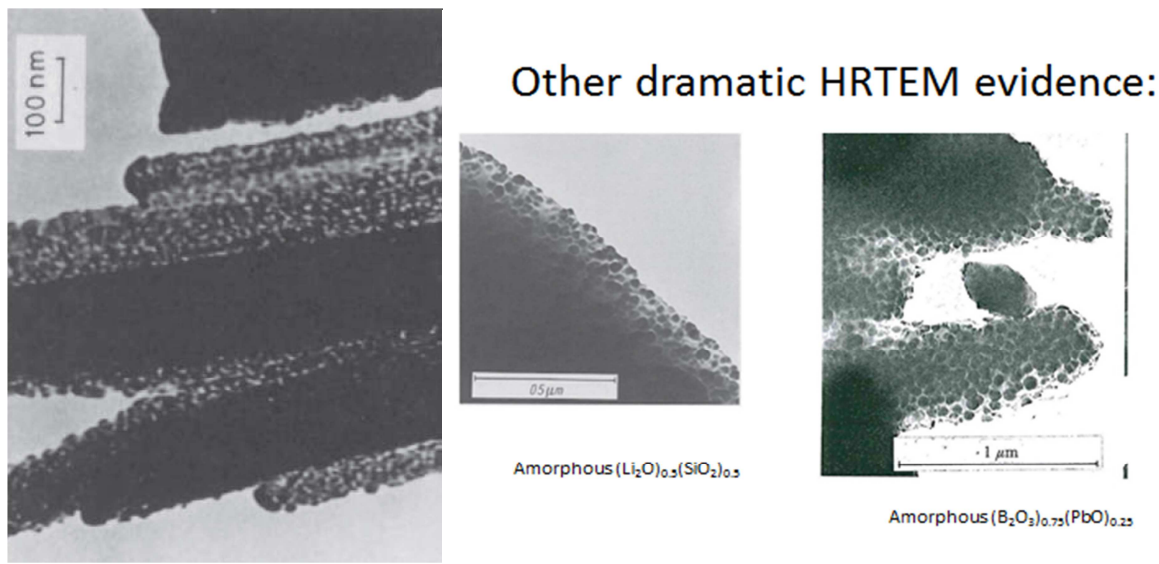

Figure 4- The intermediate-range structure of some glasses (from the left: $a$-silica, $a$-lithium-silicate, $a$-lead-borate), as seen by HRTEM. [3]

In fact, work with the ETM on low-temperature data obtained in experiments (thermal, dielectric (static and dynamic) as well as magnetic) yield indirect estimates of the RER sizes that are in some agreement (the glasses have different compositions) with the above-mentioned values [3].

## Estimate of cell sizes

| | HRTEM pictures | | Low-T physics experiments (indirect) | | |
|---|---|---|---|---|---|
| SiO$_2$ | Li$_2$O-SiO$_2$ | B$_2$O$_3$ -PbO | BAS (AlBaSiO) | Duran | BK7 |
| 500 | 600 | 500 | 163 | 154 | 300 |
| Angstrom | Angstrom | Angstrom | Angstrom | Angstrom | Angstrom |
| HRTEM | HRTEM | HRTEM | THIS LOW-T THEORY | THIS LOW-T THEORY | THIS LOW-T THEORY |

### 3. THE EXTENDED TUNNELING MODEL

We now plunge to cryogenic temperatures, below 4 K, for at such temperatures the cellular structure of glasses first revealed itself indirectly. The basic idea, given that scattering-methods, NMR-spectroscopy etc. hardly give us useful information, is to exploit the relevant "cryogenic" equilibrium degrees of freedom to gain information on the medium-range structure of glasses. It is well-known that amorphous solids (bulk glasses and amorphous films) are ridden with defects called tunneling systems (TS) and in particular (double-welled) two-level systems (2LS). Proposed in 1972 by P.W. Anderson et al. their atomic origin has never been fully clarified, but for the time being we can just associate them with local atomic defects located in the glass' structure whatever it may be (blue speckles in Fig.3, the jumbled-up lines representing the better-ordered structure of all of the RER). There is plenty of literature on the tunneling 2LS, they are usually considered as non-interacting and their presence seems to present a roadblock for the development of Josephson-junction based quantum computing devices and other quantum-information wonders. Hence the great interest in further studying them. The tunneling 2LS are associated with a local Hamiltonian

$$H_{0(2)} = -\frac{1}{2}\begin{pmatrix} \Delta & \Delta_0 \\ \Delta_0 & -\Delta \end{pmatrix} \quad (3.1)$$

with $\Delta$ called the asymmetry and $\Delta_0$ the tunneling amplitude. There is a distribution

$$P_{2LS}(\Delta, \Delta_0) = \frac{\bar{P}}{\Delta_0} \quad (3.2)$$

associated with these defects' parameters, a distribution that is convenient to use but that has never been fully justified. This couple of equations make up the so-called standard tunnelling model (STM) of tunnelling defects in glasses, which has had quite some success in explaining experiments at $T < 4$ K in the amorphous solid state.

However, in order to explain a plethora of puzzling experiments at low $T$ for bulk glasses both as a function of composition $x$ (e.g. for the series $(M_2O)_x(SiO_2)_{(1-x)}$ with M an alkali metal) and in the presence of a magnetic field $\boldsymbol{B}$ - but also in order to explain deviations from the STM's predictions in non-ideal glasses – the ETM was proposed by us in 2004. Along with the STM's 2LS, we postulated the existence of an extra set of TS that are sensitive to the magnetic field and arise from local collective phenomena. These ATS (anomalous TS) of the ETM are characterised by the Hamiltonian

$$H_{0(3)} = \begin{pmatrix} E_1 & D_0 e^{i\phi/3} & D_0 e^{-i\phi/3} \\ D_0 e^{-i\phi/3} & E_2 & D_0 e^{i\phi/3} \\ D_0 e^{i\phi/3} & D_0 e^{-i\phi/3} & E_3 \end{pmatrix} \qquad (3.3)$$

where $E_1$, $E_2$, $E_3$ are asymmetries between the three wells of a three-welled potential acting on a collective, or fictitious "particle" having collective charge $Q$, collective tunneling amplitude $D_0$ and collectively spanning some overall magnetic-field threaded surface $S_o$. $\phi = 2\pi\Phi(\boldsymbol{B})/\Phi_0$ is the Aharonov-Bohm quantum-mechanical phase picked up by the fictitious particle while tunneling the whole of the triangular path round the three-welled tunneling potential threaded by a magnetic flux $\Phi(\boldsymbol{B}) = \boldsymbol{B} \cdot \boldsymbol{S_o}$. $\Phi_0$ is the appropriate flux quantum for charge $Q$. The distribution of the tunneling parameters must be changed, since the fictitious collective quasi-particle is in the proximity of quasi-ordered RER, and reads

$$P_{ATS}(E_1, E_2, E_3; D_0) = \frac{P^*}{(E_1^2 + E_2^2 + E_3^2)D_0} \qquad (3.4)$$

where $P^*$ (dimensionless, $\approx O(1/2\pi)$), like $\bar{P}$ ($\approx 10^{45}$ J$^{-1}$), are material constants. The ETM thus considers glass as being ridden with such defects, two types (2LS+ATS), with the ATS only as the magnetic-sensitive probes. Is this real-space picture (as opposite to the popular potential- or free-energy landscape pictures) of glass realistic, and where are the two species of TS located in the proposed cellular glass structure? Well, there is mounting evidence [4] that the 2LS are placed at the boundaries between the RER where the latter would touch and coalesce were the chemical bonds placed correctly (more later), while it is clear that the ATS are collective degrees of freedom for the fluid-like particles sitting in the "voids" between the close-packed RER. Then why a three-welled ATS potential? We can come to grips with such a situation, considering that on the external surface of each RER a dense covering of $O^{2-}$ ions must exist and that each section of this surface in each "void" experiences the (screened) Coulomb potential from the opposite THREE RER surface sections, thus generating strong repulsive forces. Associating a quasi-particle to the $O^{2-}$ ions (for their number $N_{int}$ is high, $\approx$ 100 to 600) on each RER section, the simplest phenomenological model one can cook up for the quasi-particle has obvious three-fold topology (not symmetry, due to disorder). Invoking surface-chemistry adsorption mechanisms, one can envisage a temperature dependence of $N_{int}$ as reported in Fig.3

$$N_{int}(T) = N_0 \, e^{-E_0/k_B T} + N_1 \qquad (3.5)$$

with $E_0$, $N_0$ and $N_1$ ($\ll N_0$) material-dependent constants. Making use of the collective nature of the tunnelling parameters, one can quite reasonably conjecture that the three effective tunneling parameters $Q$, $D_0$ and $D = \sqrt{E_1^2 + E_2^2 + E_3^2}$ (actually, their natural cut-offs) all scale proportionally to this $N_{int}$ with temperature (the RER consolidating at the expense of the "void's" or pore's fluid-like material). Below, some pictures of the three-welled ATS tunnelling potential and its spectrum [1,3].

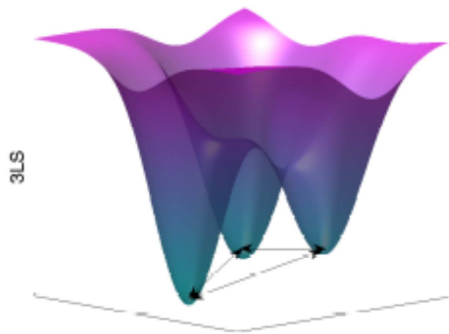
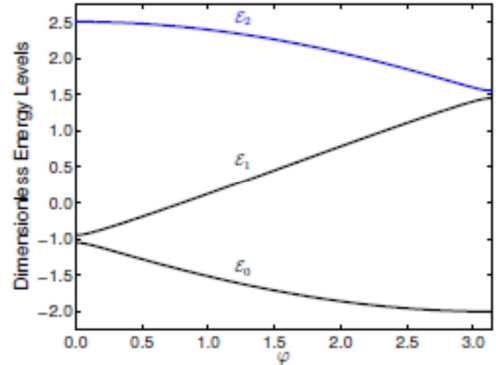

The particle moving in the three-welled potential.

The use of the above ETM has lead us to the consistent (the model and its parameters are always the same!) explanation of a plethora of experiments at low- and very-low temperatures, without any need to take the 2LS or the ATS as interacting entities (the ATS are in fact quasi-particles of already strongly-interacting groups of atomic particles, anyway). Below, we re-propose some of the best results: for the heat-capacity, the dielectric constant and the polarization echo amplitude [1,3]. They all show the same qualitative dependence on the magnetic field $B$ and this is coming from the magnetic-field dependence of the density of states (DOS) of the ETM model. Much remains, however, to be understood at the lowest temperatures, though they are only (yet very important) details here.

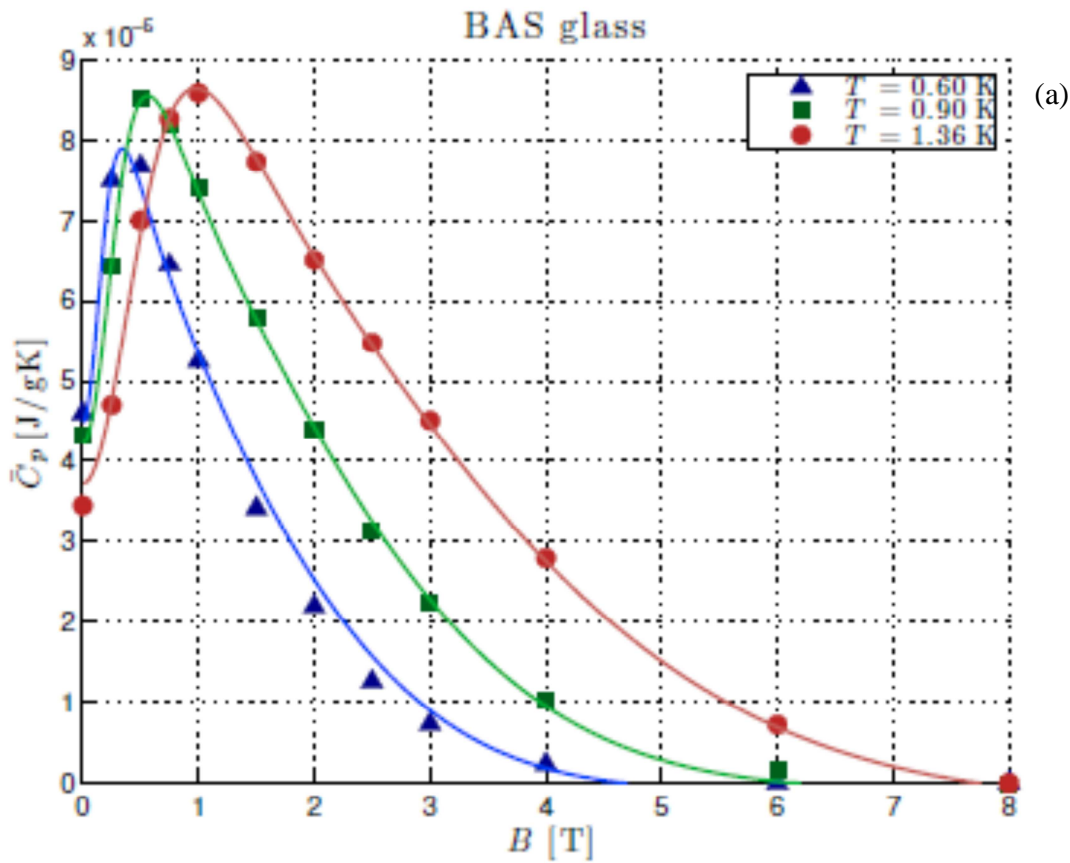

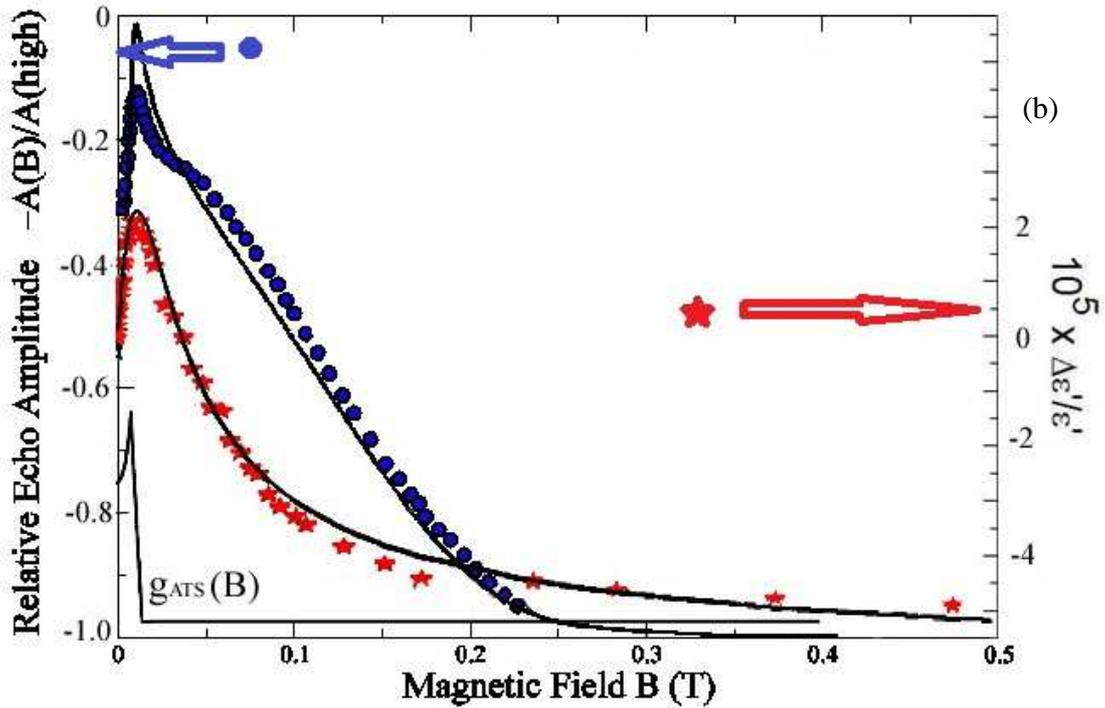

Figure 5- The ETM at work (continuous lines) for the theoretical interpretation of heat-capacity($C_p$) data (a) and then for the dielectric constant (real part, $\varepsilon'$, red points) and polarization-echo amplitude ($A$, blue points, sign changed!) data (b) for BAS-glass in the presence of a magnetic field $B$. All curves derive from convolutions of the ATS density of states $g_{ATS}(B)$ which is also drawn. [3]

## 4. THE ETM AND THE BOSON PEAK

Very briefly, since work is still in its infancy with this approach. The idea of a cellular structure of glasses to explain the cryogenic anomalies was in fact first proposed by H.P. Baltes, working in the early 1970s. The Boson-Peak problem consists of a broad peak in $C_p/T^3$ (see below) as well as two less-talked-about issues: the excess heat-capacity for the amorphous versus the crystalline Debye-Einstein formula, and the peak in $g(\omega)/\omega^2$ ($g(\omega)$ being the vibrational DOS) in the terahertz region. In fact, all three issues should be described by the same vitreous-state scenario, save for the peak in $g(\omega)/\omega^2$ which is not exclusive of glasses (being present, for example, also for quartz (crystalline $SiO_2$)). Baltes demonstrated that the excess heat-capacity and the linear temperature-dependence contributions to $C_p$ were consequences of the presumed cellular structure of glasses, however the linear $T$-dependent contribution to $C_p$ together with the logarithmic one for the excess $\varepsilon'$ (not described by the cellular model of Baltes) at cryogenic temperatures were then resolved with the affirmation of the 2LS STM. In our present approach, the cellular model is completed by the existence of a fluid-like medium in which the RER are sitting and therefore by the 2LS+ATS picture (Fig.3, amended by a proper re-positioning of the (blue) 2LS) which can be taken to higher temperatures (in fact all the way to $T_g$) by suitably taking the associated forces into account.

Then, the problem of the vibrational spectrum of glasses at intermediate temperatures can be attacked by studying the system pictured below – closed-packed rigid cells immersed in a fluid-like medium – and then turning the resulting $g(\omega)$ into a heat-capacity calculation. This research topic is still being developed.

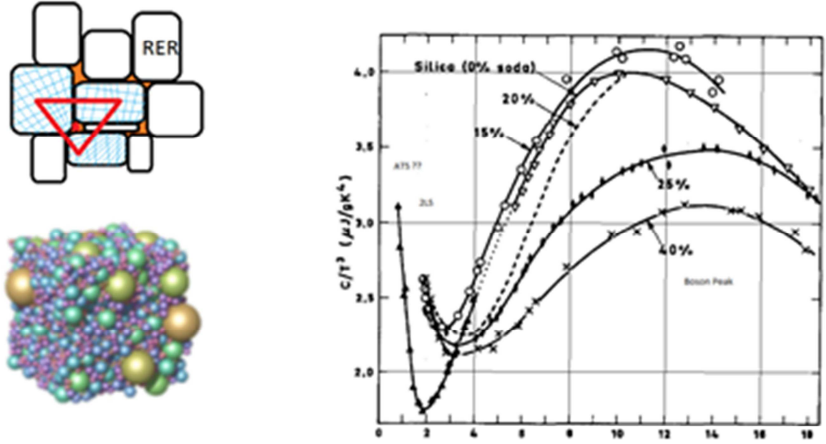

Figure 6 – The main issue of the Boson-Peak behavior (to the right in the figure) as well as the peak in $g(\omega)/\omega^2$ and excess heat-capacity should be all consequences of the cellular structure of glasses too, with the nature of the forces keeping the cellular mass of the glass mechanically stable deriving from the modeling in terms of 2LS and ATS effective degrees of freedom. [4]

## 5. THE ETM AND THE GLASS TRANSITION

Probably this is the crux of this paper, for most interested people attending this Symposium. Can the cellular-model, completed with the ETM formulation, be taken all the way to $T_g$ and above, in the super-cooled state of a glass-forming liquid? Let us begin with a very simple argumentation: how to derive the quench-rate, $\kappa$ (dimensions, e.g., K/s and often denoted also by $R$), dependence of the putative glass "transition" temperature $T_g=T_g(\kappa)$. We can imagine there is a dominant relaxation time $\tau_o$, so that a way to determine the configurational entropy $S_c$ per particle is to write ($T_m$-$T$ being the temperature interval spanned by fast cooling from the melting temperature $T_m$ to the current temperature $T$):

$$S_{conf} = k_B \ln\left(\frac{T_m - T}{\kappa \tau_o}\right) \qquad (5.1)$$

(the fraction in brackets being the number of ways one can divide the $[T,T_m]$ interval in temperature ticks $\kappa\tau_o$). According to the coherently-rearranging clusters (CRC) approach of Adam and Gibbs (AG), the above could also be evaluated, on dimensional grounds, as

$$S_{AG} = \frac{E_0}{k_B T} \quad \rightarrow \quad \frac{E_0}{k_B(T-\theta)} \qquad (5.2)$$

Here, the first expression would refer to uncorrelated AG's CRC and the expression on its right should be the corresponding "molecular field" correction which would take into account the cooperativity, via short-range CRC interactions, of the AG clusters (our own RER) in the temperature region below $T_{MC}$. Indeed, the entropy density is the best candidate for an order parameter's susceptibility for the glass-transition problem. The "Curie-Weiss" temperature $\theta$ could, of course, be understood as the Kauzmann temperature $T_K$. Combining the two expressions Eq.s (5.1) and (5.2) gives the temperature (in fact the temperature range) where the two expressions are consistent. One finds, indeed, that there are two delimiting temperatures (see graph below) which can be identified with the putative $T_g$ and the mode-coupling $T_{MC}$, respectively [4]. Detailed inspection gives (due to the logarithm in Eq. (5.1)) a logarithmic dependence of $T_g$ on $\kappa$ (or $R$) as is observed in both experiments and simulations. But also a limiting value $R_{max}$ of $R$ (or $\kappa_{max}$ of $\kappa$) at which $T_g = T_{MC}$ and above which (for $R > R_{max}$) vitrification can no longer take place in a finite temperature-interval (an experimentally observable, important fact and a theoretically interesting prediction indeed [4] if confirmed: it could be interpreted by saying that for $R > R_{max}$ a true second-order phase transition sets in).

## Location of Glass Transition as a fn. of Cooling Rate R

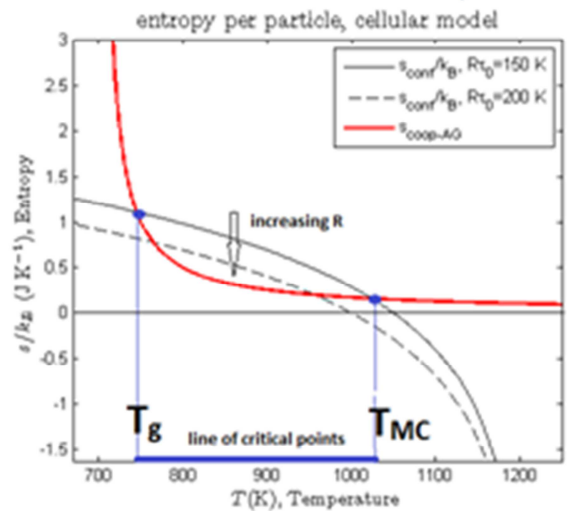

$$s_{conf} = k_B \ln \frac{T_m - T}{\kappa \tau_0}$$

$$s_{AG} = \frac{E_0}{k_B T} \rightarrow \frac{E_0}{k_B (T - \Theta)}$$

much like a «dynamical» molecular field theory for RER-RER interactions

($\Theta$ Curie-Weiss like temperature, here Kauzmann temperature)

For $(Na_2O)_x(SiO_2)_{1-x}$ one finds $T_g \approx 750$ K

and this explains why the Glass «Transition» is truly only a Crossover (gradual jamming of RERs)

So we now have, from the above naïve argument, a line of temperatures in the interval ($T_g$, $T_{MC}$) where the RER should be highly correlated and something like **RER-jamming** should occur.

## Glass «Transition»=Gradual Jamming

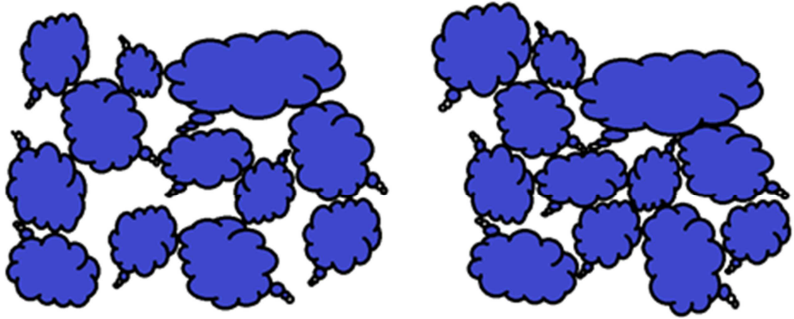

$T_{MC}$ > T > $T_g$

The above cartoon shows how this may occur. The RER are represented as rough clusters that begin to touch each other, perhaps precisely at temperature $T_{MC}$, after their growth below $T_x$, and re-arrange themselves like grains of a shaked powder (yet definitely still growing) as the lower critical temperature (something like $T_g$ or even $T_K$, so then θ would be a new characteristic temperature) is approached. The rougher and more asymmetric the RER cells, the wider the *[$T_g$, $T_{MC}$]* interval, and indeed a strong glass like *a*-SiO$_2$ would have a short jamming interval, while an organic or, worse, polymeric glass would have a wider glass-"transformation" interval. This is indeed often observed (e.g. in glycerol, having rather elongated molecules).

How to make, then, this qualitative picture more precise? Well, it does look like a Kosterlitz-Thouless-Halperin-Nelson-Young (KTHNY) 2D-melting scenario (developed in the '80s for a 2D crystal) and this could be its first manifestation in a 3D situation. Starting from the cellular model we can investigate how – by warming up glass in a succession of quasi-static, equilibrium thermodynamic states (a reversible transformation) – the glass may melt. We must in fact argue, in truth at this point, that the "melting" of glass should be a completely different type of transformation than the reverse, glass-"formation" (no transition) process. In fact, we can imagine obtaining the super-cooled liquid (and then the classic liquid) by slowly (quasi-statically) warming up the glass, while we must absolutely make sure – on the other hand - to quickly quench the liquid (a totally non-equilibrium, irreversible procedure) in order to regain the glass. Were we to slowly (quasi-statically) cool down the just-melted glass-forming substance again, we would obtain a crystal (in fact a poly-crystalline solid typically) and no glass at all. This is why it is hard to believe in the existence of a second-order "equilibrium" glass transition even for the "ideal glass". But we can melt the glass slowly, through a succession of quasi-equilibrium states, and thus use equilibrium statistical mechanics to describe the super-cooled liquid (and then the ordinary liquid) we obtain in this way.

Let us consider the cellular glassy medium (well below $T_K$, say) and try to figure out how, by thermally exciting "defects", we may break-up the glass into a super-cooled liquid as described in Fig.s 1 and 2. Clearly, this is obtained by creating (Fig. 7) pairs of infinitely-long "voids" (still filled with fluid-like material, like it exists in-between the close-packed RER) in the shape of (distorted, in practice) **defect-cones** with their vertices a distance $r_{12}$ apart. It is hard to do the full calculation still, but by calculating (on the basis of the ETM-generated force fields associated with the proposed 2LS+ATS distribution) the energy change *ΔE(L)* for a single defect-cone of length *L* inserted in the perfectly close-packed cellular glassy solid, we may then infer what the form of the defect-cone – defect-cone interaction should be.

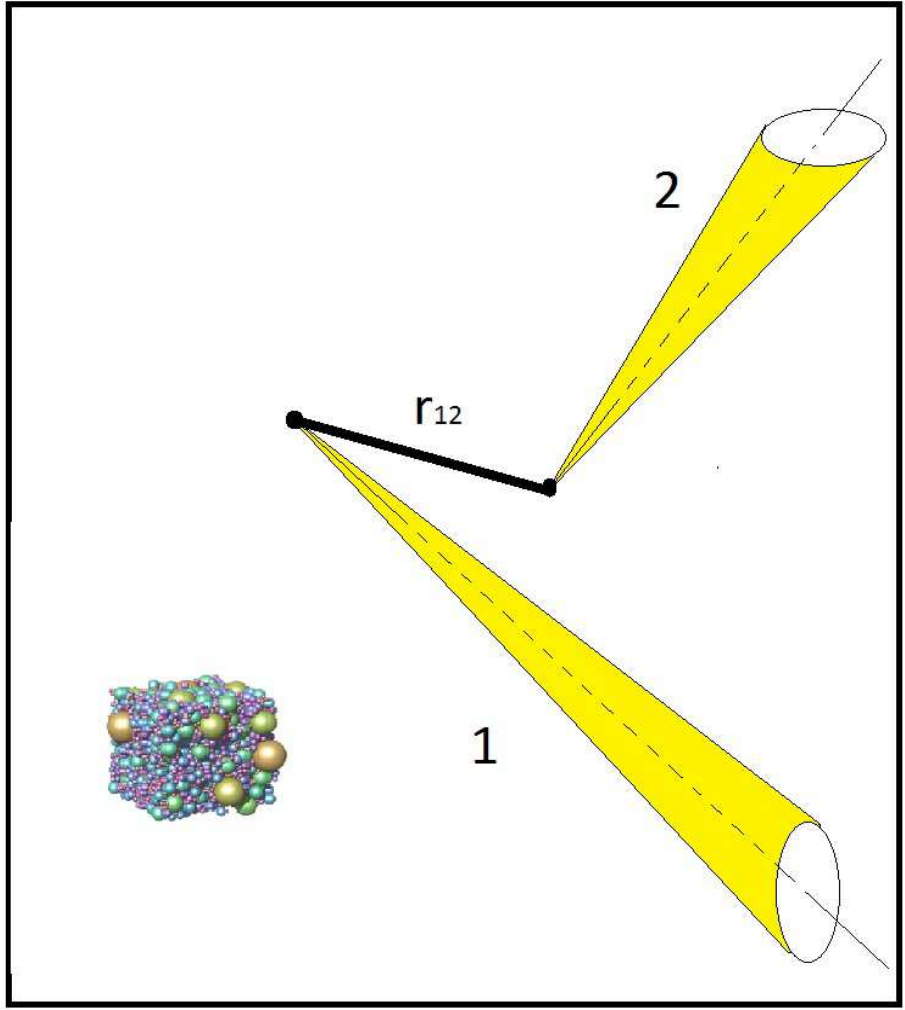

Figure7- A cartoon of the cellular glass (3D closed-packed RER structure in the inset, which should actually be permeating all of the space except the two cones) and schematic representation of an unbound pair (yellow cones) of "voids" in the cellular glass, the defect-cones actually being filled with liquid-like atomic/molecular material. The proliferation of such unbound pairs for T > $T_0$ clearly breaks up a solid aggregate like in the inset and produces the DH super-cooled liquid. [4]

Working out the maths for a single perfect cone, for simplicity, the 2LS+ATS cellular scenario leads to [4]:

$$\Delta E(L) = k_o \ln(\frac{S_{VL}}{s_0}) \qquad (5.3)$$

where $S_{VL}$ is the lateral surface of the defect-cone (or defect-line, in general a "vortex" line (VL)) of length $L$ and $k_o$, $s_0$ are some geometric and material constants. This result follows from the summation of all the attractive (2LS) and repulsive (ATS) energy contributions inside the cone, where in fact what matters is the energy change in the lateral defect-cone surface. A kind of dimensional reduction from 3D to 2D. Since for a cone $S_{VL}=\frac{2}{3}\pi L^2 \tan \alpha$ (where $2\alpha$ is the cone's aperture) we see that $\Delta E(L)$ growths like $\ln(L/r_0)$ and we infer that the energy cost of inserting a pair of defect-cones with their vertices a distance $r_{12}$ apart is given by

$$\Delta E(r_{12}) = c_o \ln\left(\frac{r_{12}}{r_0}\right) \tag{5.4}$$

where $r_0$, $c_0$ are new constants dependent on the angles of aperture and relative angular orientation of the two defect-cones in a pair. At this point, we consider a dilute gas of bound defect-cones distributed in the cellular-structured glass; suppose there are $N$ such pairs of defect-cones per unit volume, so the energy per unit volume of this dilute gas of pairs of defect-cones is

$$E_{DGas} = NE_0 - Nc_0 \ln\frac{N^{-\frac{1}{3}}}{r_0} = NE_0' + \frac{1}{3}N\, c_0 \ln N \tag{5.5}$$

(the negative sign denoting defect-cone – defect-cone attraction, $E_o$ etc. being core-energy constants). Assuming equilibrium thermodynamics, we can add the entropy term of defect-cone bound-pair location to get the free-energy per unit volume

$$\begin{aligned}F_{DGas} &= NE_0'' + \frac{1}{3}N\, c_0 \ln N - Tk_B \ln N! + const. \\ &= NE_0''' + \frac{1}{3}N\, c_0 \ln N - Nk_B T \ln N\end{aligned} \tag{5.6}$$

($E_0'''$ now containing a term linear in $T$). Now we can see that for $T < T_0$ (with $T_0 = \frac{1}{3}c_0/k_B$) we get an increase of free-energy, meaning that the defect-cones pairs cannot dissociate, while for $T > T_0$ dissociation takes place. The defect-cone pairs' density at equilibrium can be determined by minimizing Eq. (5.6) and the result is ($E_{oo}$ being an overall positive core-energy term, $n_o$ the value at $T=\infty$)

$$N_o(T) = n_o e^{-\frac{E_{oo}}{k_B(T-T_0)}} \tag{5.7}$$

So there is a density of defect-cone pairs above $T_0$ (now to be identified with Kauzmann's $T_K$) which vanishes in a highly singular way near $T_K$. If a tiny ball is made to move in this super-cooled liquid filled with unbound defect-cone pairs, it will experience a viscosity given by the VFT-law:

$$\eta \propto N_o(T)^{-1} \propto e^{\frac{B}{T-T_K}} \tag{5.8}$$

This is of course the "naïve" Thouless argument, as was envisaged by P.W. Anderson [5] without any knowledge of what the unbinding "defects" might be; the cellular and ETM scenarios can now provide this knowledge. More precise Kosterlitz-type renormalization group calculations shall provide more precise temperature-dependent forms for $\eta = \eta(T)$ to be tested in fitting viscosity and other transport measurements in real glasses.

The final job of destroying the super-cooled liquid state will be done by the disclinations' unbinding, much as in KTHNY-theory of 2D melting, and the obvious choice for the disclinations in a slow-particle (and better-ordered) cluster are the disclinations characterizing the putative crystalline counterpart(s) of the solid that never formed. These should unbind at $T_{MC}$.

## 6. CONCLUSIONS

We have seen how, by starting from the low-temperature (cryogenic temperatures) end and then warming up all the way to $T_g$ and beyond, a comprehensive theory of the vitreous state begins to take shape, providing a coherent description of the main aspects and phenomena of the physics of glasses. Beside the cryogenic region (where many more problems are still open, but the theory has been already developed at length) only a few touch-and-go aspects of the higher temperatures' issues have been tackled so far. The theory, for these reasons, is still in the making and especially towards the kilo-K end. However, while many tassels of the puzzle of glass physics are still missing, the general picture is definitely beginning to emerge.

# 7. STOP PRESS – VERY RECENT RESULTS: DIRECT EXPERIMENTAL EVIDENCE FOR THE CELLULAR EXTENDED-TUNNELING MODEL

We have recently performed some remarkable experiments in order to give strength to the above-proposed picture of glasses and glass-forming liquids. The results fully confirm, once again but now tangibly, the above proposed vitreous-state scenario completely.

The first experiment consists in measuring the (tiny) magnetization $M$ of a nominally non-magnetic multi-component silicate glass (e.g.) in a SQUID-magnetometer. The contribution of paramagnetic impurities - overwhelmingly Fe for the silicates - must be taken into account and this is done by determining the Fe-group trace elements' concentration in a mass-spectrometer. Subtracting the paramagnetic-Fe Langevin (as well as the bulk diamagnetic Larmor) contributions, one is left with a **non-zero *intrinsic* magnetization** due to the glass *per se*. The intrinsic contribution from the glass (in fact due to coherent tunnelling currents [6] within the "voids" between the RER) has a completely new, non-monotonic $T$- and $B$- (or $H$-, where $B=\mu_0 H$) dependence. Fig.8 shows some of the experimental data [7] for some glass samples we investigated, $M=M(T,H)$ as a function of $T$. The non-monotonous behaviour is well explained by the ETM model illustrated in Section 3, with basically the same material parameters as for the other low-$T$ experiments.

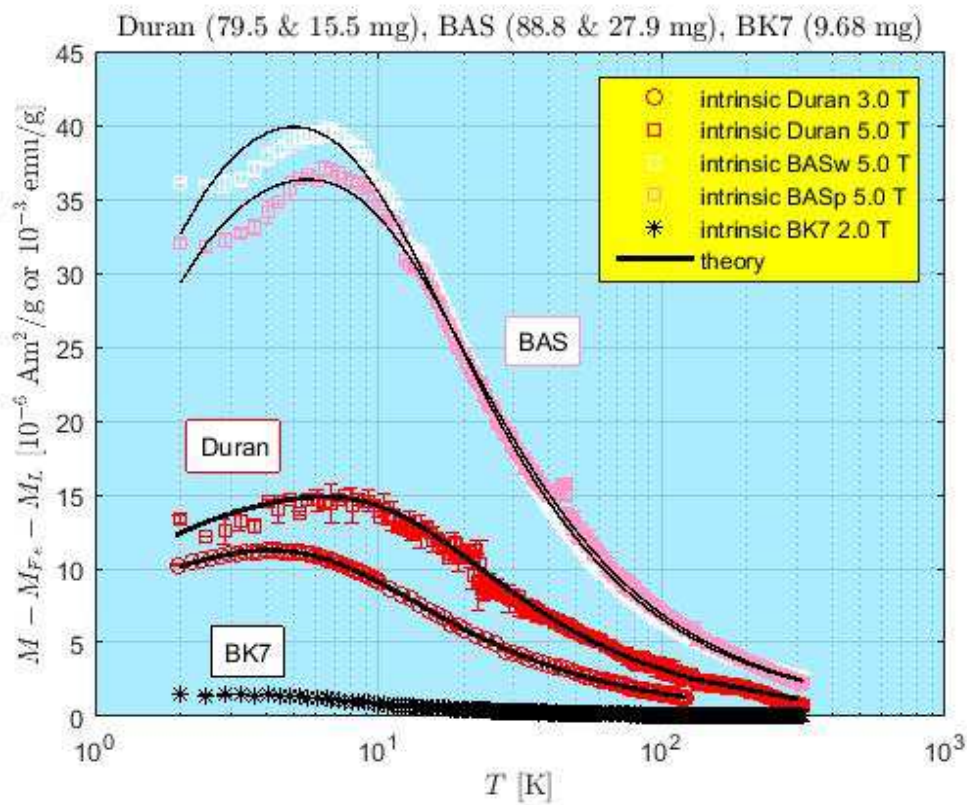

Figure 8 – The intrinsic glass magnetization measured for three multi-component silicates; theory is the ETM [6,7].

The second (morphological) experiment was the fabrication of a bulk glass specimen made from a multi-component glass-forming material where some poly-disperse seeding powder has been added to the molten liquid. In place of inducing an "ideal glass", the procedure has induced - instead of a clear and homogeneous vitreous mass as expected - a cellular structure that is **the sub-millimetric version of the proposed nanometric cellular model**. Many other features of the ETM have also been verified in this solid specimen, making the cellular- and ETM-scenario apparent and **visible to the naked eye**. Fig.9 shows some pictures that were obtained by us through ordinary means.

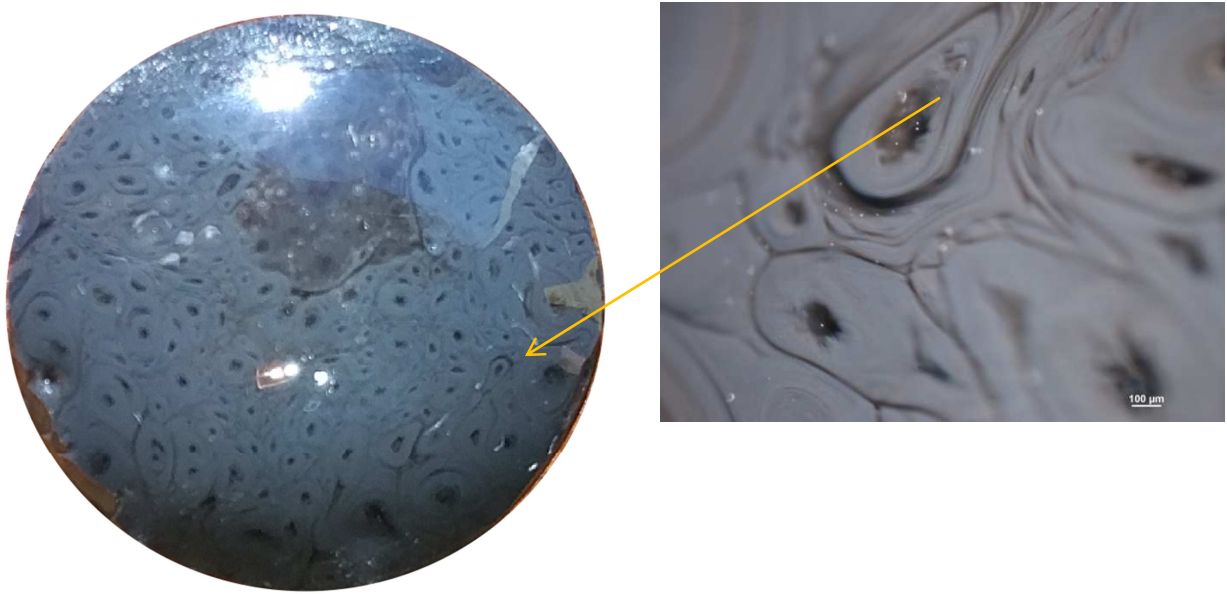

Figure 9 – Pictures (photocamera at left, optical-microscopy detail (upside down!) at right) of a glass specimen we made showing the cellular structure almost at the naked-eye scale. The black seeding speckles ignored, these pictures are the sub-millimetric version of the usually nanometric structure of ordinary glasses as proposed and advocated in this work. Compare these pictures with Fig.s 3 and 4 in this work. [7]

## ACKNOWLEDGEMENTS

The Author acknowledges support by INFN through project Iniziativa Specifica GEOSYM-QFT.